\documentclass[10pt,journal,a4paper]{IEEEtran}
\usepackage{amssymb}
\usepackage{amsmath}
\usepackage{bm}
\usepackage{amsthm}
\usepackage{graphicx}
\usepackage{subfigure}
\usepackage{cite}
\usepackage{enumerate}
\usepackage{mathrsfs}
\usepackage{multirow}
\usepackage{amsfonts}
\usepackage{url}
\usepackage{clrscode}
\usepackage{amsthm}
\usepackage{CJK}
\usepackage{indentfirst}
\usepackage[ruled]{algorithm2e}
\usepackage{algorithmic}

\allowdisplaybreaks[3]

\begin{document}
\title{Cellular Controlled Cooperative Unmanned Aerial Vehicle Networks with Sense-and-Send Protocol}

\author{
\IEEEauthorblockN{\normalsize{Shuhang Zhang, Hongliang Zhang, {\em{Student Member}}, {\emph{IEEE}}, Boya Di, {\em{Student Member}}, {\emph{IEEE}},\\ and Lingyang Song, {\em{Senior Member}}, {\emph{IEEE}}}}\\
\thanks{The authors are with the School of Electronics Engineering and Computer Science, Peking University, Beijing, China. Email:$\{$shuhangzhang, hongliang.zhang, diboya, lingyang.song$\}$@pku.edu.cn.}
}
\maketitle

\begin{abstract}
In this paper, we consider a cellular controlled unmanned aerial vehicle~(UAV) sensing network in which multiple UAVs cooperatively complete each sensing task. We first propose a sense-and-send protocol where the UAVs collect sensory data of the tasks and transmit the collected data to the base station. We then formulate a joint trajectory, sensing location, and UAV scheduling optimization problem that minimizes the completion time for all the sensing tasks in the network. To solve this NP-hard problem efficiently, we decouple it into three sub-problems: trajectory optimization, sensing location optimization, and UAV scheduling. An iterative trajectory, sensing, and scheduling optimization (ITSSO) algorithm is proposed to solve these sub-problems jointly. The convergence and complexity of the ITSSO algorithm, together with the system performance are analysed. Simulation results show that the proposed ITSSO algorithm saves the task completion time by 15\% compared to the non-cooperative scheme.
\end{abstract}

\begin{IEEEkeywords}
unmanned aerial vehicle, sense-and-send, trajectory optimization, cooperative sensing.
\end{IEEEkeywords}
\section{Introduction}
Unmanned aerial vehicle (UAV) is an emerging facility which has been widely applied in military, public, and civil applications \cite{ENCA2017}. Among these applications, the use of UAV to perform data sensing has been of particular interest owing to its advantages of on-demand flexible deployment, larger service coverage compared with the conventional fixed sensor nodes, and additional design degrees of freedom by exploiting its high mobility~\cite{WJHRMH2017, MTA2016,ZZHBS2018}. Recently, UAVs with cameras or sensors have entered the daily lives to execute various sensing tasks, e.g. air quality index monitoring~\cite{YZBSH2018}, autonomous target detection~\cite{KGM2017}, precision agriculture~\cite{AMCG2017}, and water stress quantification~\cite{SDWC2017}. Besides, the sensory data collected in such tasks needs to be immediately transmitted to the base stations~(BSs) for further processing in the servers, thereby posing low latency requirement on the wireless network.

To this end, the cellular network controlled UAV transmission is considered to play an important role in satisfying the low latency requirement. In the traditional UAV ad hoc sensing network~\cite{JCZYRL2017,TBKS2018}, the sensory data is transmitted through UAV-to-UAV and UAV-to-ground communications over unlicensed band, which cannot guarantee the QoS requirements. Recently, the network controlled UAVs are proposed to transmit sensory data to BSs directly through the cellular network~\cite{BAS-2016,NMT2017,ZZDS2018}. In~\cite{BAS-2016}, the authors analyzed the use of LTE for realizing UAV sensing network, which improves the data rate. In~\cite{NMT2017}, the potential of UAVs as Internet of Things devices was discussed, which can reduce the latency of the network. In~\cite{ZZDS2018}, a cellular UAV sensing network was proposed to improve the data rate.

In this paper, we study a single cell\footnote{The multiple cell scenario is an extension of the single cell scenario, and will be studied in the future.} UAV sensing network in which UAVs perform sensing tasks and transmit the collected data to a BS. Note that sensing failure may occur due to imperfect sensing in practical systems. Therefore, we advocate UAV cooperation in the sensing networks to further improve the successful sensing probability~\cite{HMGIYR2015}. To be specific, multiple UAVs are arranged to collect the sensory data for the same sensing task and to transmit the collected data to the BS separately. In this way, the successful sensing probability requirement for each UAV is loosened~\cite{HAE2013,CLHSC2010}, and the task completion time of each UAV can be shortened~\cite{SSRWRSN2015}.

Although UAV cooperation has the advantages in reducing the sensing failure probability and task completion time, it also involves some challenges. Firstly, as the UAV scheduling will influence the sensing performance, an efficient scheduling scheme is necessary. Secondly, since each sensing task is performed by multiple UAVs, the trajectories and sensing locations of UAVs are coupled with each other. In light of these issues, we first propose a sense-and-send protocol to support the cooperation and facilitate the scheduling. Then, we optimize the trajectories, sensing locations, and scheduling of these cooperative UAVs to minimize the completion time for all the tasks. As the problem is NP-hard, we decompose it into three subproblems, i.e., trajectory optimization, sensing location optimization, and UAV scheduling, and solve it by an iterative algorithm with low complexity.

Note that in literature, most works focused on either sensing or transmission in UAV networks, instead of joint considering UAV sensing and transmission. The work in~\cite{IKMA2010} presented a platform to deal with the cooperation and control of multiple UAVs with sensing and actuation capabilities for load transportation and deployment. The estimation problem for both the position and velocity of a ground moving target was addressed in~\cite{GPCAM2006} using a team of cooperative sensing UAVs. In~\cite{WZRPRL2017}, a searching algorithm to make multiple UAVs autonomously and cooperatively search roads in the urban environments was proposed. In~\cite{WZZ2018}, the authors considered a multi-UAV enabled wireless communication system, in which the UAVs work as BSs cooperatively to serve the ground users taking the fairness into consideration. Multiple UAVs were deployed as wireless BSs to provide a better communication coverage for ground users in~\cite{MSBD2016}.

The main contributions of this paper can be summarized below.
\begin{enumerate}[(1)]
\item We propose a cooperative UAV sensing network where multiple UAVs complete the same sensing task cooperatively, and transmit the collected data to a BS. A sense-and-send protocol is designed to support such cooperation.

\item We decompose the joint optimization problem into three sub-problems: trajectory optimization, sensing location optimization, and UAV scheduling, and propose an iterative trajectory, sensing, and scheduling optimization~(ITSSO) algorithm to jointly solve the sub-problems. The theoretical performance analysis is then studied.

\item Simulation results verify the theoretical analysis, and show that the proposed cooperative sensing scheme outperforms the non-cooperative and fixed sensing location schemes.
\end{enumerate}

The rest of this paper is organized as follows. In Section~\uppercase\expandafter{\romannumeral2}, we describe the system model of the cooperative UAV sensing network. In Section~\uppercase\expandafter{\romannumeral3}, we elaborate the sense-and-send protocol for the cooperative UAVs. In Section~\uppercase\expandafter{\romannumeral4}, we formulate the task completion time minimization problem by optimizing the trajectory, sensing location, and UAV scheduling. The ITSSO algorithm and the algorithm analysis are proposed in Section~\uppercase\expandafter{\romannumeral5}. Simulation results are presented in Section~\uppercase\expandafter{\romannumeral6}, and finally we conclude the paper in Section \uppercase\expandafter{\romannumeral7}.
\section{System Model}
\begin{figure}[h]
\centering
\includegraphics[width=3.2in]{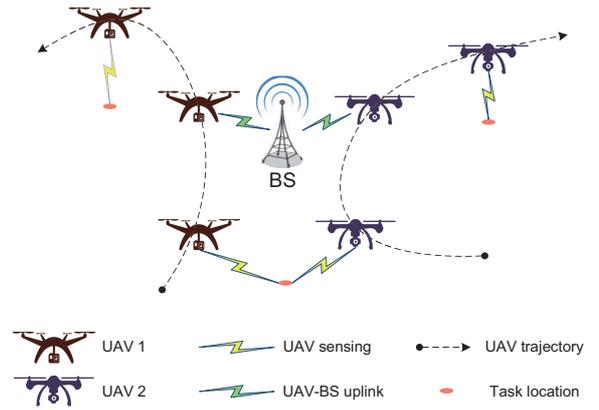}
\caption{System model for UAV cooperation.}
\label{scenario}
\end{figure}

We consider a single cell OFDM cellular network as shown in Fig.~\ref{scenario}, which consists of one BS, $M$ UAVs, denoted by ${\cal{M}}=\{ 1, 2, ..., M\}$, and $K$ orthogonal subcarriers, denoted by ${\cal{K}}=\{ 1, 2, ..., K\}$. Within the cell coverage, there are $N$ sensing tasks to be completed, denoted by ${\cal{N}}=\{ 1, 2, ... N\}$. The UAVs perform each task in two steps: UAV sensing and UAV transmission, and thus, these two procedures will be repeated by the UAVs until all the tasks are completed. Note that different types of sensing tasks require UAVs with different sensors~\cite{SDO2018,VKFNK2017}. Therefore, the sensory data of each task is collected by a predefined UAV group cooperatively, and the UAVs in this group send the collected data to the BS separately\footnote{Unlike most of the previous works, which typically treat UAVs as relays or BSs~\cite{HZZZ2018,ZZL2016}, our work considers the UAV as a flying mobile terminal in the UAV sensing network.}. We denote the UAV group that performs task $j$ by $\mathcal{W}_j$, satisfying $\mathcal{W}_j\subseteq \cal{M}$, and $|\mathcal{W}_j|=q$. For UAV $i$, it is required to execute a subset of tasks in sequence, denoted by ${\cal{N}}_i=\{ 1, 2, ... N_i\}, \forall i \in \cal{M}$, with ${\cal{N}}_i\subseteq {\cal{N}}$\footnote{Tasks such as geological detection can be performed with this model, where each UAV is arranged to perform a series of tasks, and the geological information of each task is sensed by multiple UAVs.}. In the following, we first describe the UAV sensing and UAV transmission steps, and then introduce the task completion time of the UAVs in the network.
\subsection{UAV Sensing}
In UAV sensing, it is important to design the trajectory of each UAV along which it moves towards the locations of a sequence of tasks. Without loss of generality, we denote the location of the BS by $(0,0,H)$, and the location of task $n$ by $(x^n,y^n,0)$. In time slot $t$, let $\bm{l}_i(t)=(x_i(t),y_i(t),z_i(t))$ be the location of UAV $i$, and $\bm{v}_i(t)=(v_i^x(t),v_i^y(t),v_i^z(t))$ be its velocity, with $\bm{v}_i(t)=\bm{l}_i(t)-\bm{l}_i(t-1)$. Due to the space and mechanical limitations, the UAVs have a maximum velocity $v_{max}$. For safety consideration, we also assume that the altitude of the UAVs in this network should be no less than a minimum threshold $h_{min}$.

In time slot $t$, the distance between UAV $i$ and the BS is expressed as
\begin{equation}\label{BS distance}
d_{i,BS}(t)=\sqrt{(x_i(t))^2+(y_i(t))^2+(z_i(t)-H)^2}.
\end{equation}
The distance between UAV $i$ and task $j$ is given by
\begin{equation}\label{task distance}
d_{i,j}(t)=\sqrt{(x_i(t)-x^n)^2+(y_i(t)-y^n)^2+(z_i(t))^2}.
\end{equation}

We utilize the probabilistic sensing model as introduced in~\cite{SK2017,CRCG2013,HCB2012}, where the successful sensing probability is an exponential function of the distance between the sensing UAV and the task location. The successful sensing probability for UAV $i$ to perform sensing task $j$ can be shown as
\begin{equation}\label{UAV sensing probability}
PR(i,j)=e^{-\lambda d_{i,j}(t)},
\end{equation}
where $\lambda$ is a parameter evaluating the sensing performance. The probability that task $j$ is successfully completed can be expressed by
\begin{equation}\label{cooperate sensing}
PR_j=1-\prod_{i\in \mathcal{W}_j} (1-PR(i,j)).
\end{equation}
Define $PR_{th}$ as the probability threshold, and task $j$ can be considered to be completed when $PR_j\geq PR_{th}$.
\subsection{UAV Transmission}\label{T model}
In UAV transmission, the UAVs transmit the sensory data to the BS over orthogonal subcarriers to avoid severe interference.
We adopt the 3GPP channel model for evaluating the urban macro cellular support for UAVs~\cite{3GPPR17}.

Let $P_U$ be the transmit power of each UAV. The received power at the BS from UAV $i$ in time slot $t$ can then be expressed as
\begin{equation}\label{BS receive power}
P_{i,BS}(t)=\frac{P_U}{10^{PL_{a,i}(t)/10}},
\end{equation}
where $PL_{a,i}(t)$ is the average air-to-ground pathloss, defined by $PL_{a,i}(t)=P_{L,i}(t)\times PL_{L,i}(t)+P_{N,i}(t)\times PL_{N,i}(t)$. Here, $PL_{L,i}(t)$ and $PL_{N,i}(t)$ are the line-of-sight (LoS) and non-line-of-sight (NLoS) pathloss from UAV $i$ to the BS, with $PL_{L,i}(t)=28+22\times\log(d_{i,BS}(t))+20\times \log(f_c)$, and $PL_{N,i}(t)=-17.5+(46-7\times\log (z_i(t)))\times \log(d_{i,BS}(t))+20\times \log(\frac{40\pi\times f_c}{3})$, respectively, where $f_c$ is the carrier frequency. $P_{L,i}(t)$ and $P_{N,i}(t)$ are the probability of LoS and NLoS, respectively, with $P_{N,i}(t)=1-P_{L,i}(t)$. The expression of LoS probability is given by
\begin{equation}\label{Los}
P_{L,i}(t)\hspace{-1mm}=\hspace{-1mm}\left\{
             \begin{array}{lr}
             1, \hspace{-1mm}& \hspace{-1mm} d_{i}^{H}(t)\hspace{-1mm}\leq \hspace{-1mm}d_1,\\
             \frac{d_1}{d_{i}^{H}(t)}+e^{\left(\frac{-d_{i}^{H}(t)}{p_0}\right)\left(1-\frac{d_1}{d_{i}^{H}(t)}\right)}, \hspace{-1mm}& \hspace{-1mm} d_{i}^{H}(t)\hspace{-1mm}>\hspace{-1mm} d_1,
             \end{array}
\right.
\end{equation}
where $p_0=4300\times\log(z_i(t))-3800$, $d_1=\max\{460\times\log(z_i(t))-700, 18\}$, and $d_{i}^{H}(t)=\sqrt{(x_i(t))^2+(y_i(t))^2}$. Note that the cooperative sensing and transmission process is completed in a long time period, and small scale fading is neglected in the transmission channel model in problem~(\ref{system_optimization}).

Therefore, the signal-to-noise ratio (SNR) from UAV $i$ to the BS is given by
\begin{equation}\label{SNR}
\gamma_{i}(t)=\frac{P_{i,BS}(t)}{\sigma^2},
\end{equation}
where $\sigma^2$ is the variance of additive white Gaussian noise with zero mean. For fairness consideration, each UAV can be assigned to at most one subcarrier. We define a binary UAV scheduling variable $\psi_{i}(t)$ for UAV $i$ in time slot $t$, where
\begin{equation}
\psi_{i}(t)=\left\{
            \begin{array}{lr}
            1, \text{UAV $i$ is paired with a subcarrier},\\
            0, \text{otherwise}.
            \end{array}
\right.
\end{equation}
Therefore, the data rate from UAV $i$ to the BS is given by
\begin{equation}\label{rate}
R_{i}(t)=\psi_{i}(t)\times W_B\log_2\hspace{-1mm}\left(1+\gamma_{i}(t)\right),
\end{equation}
where $W_B$ is the bandwidth of a subcarrier.
\subsection{Task Completion Time}
For UAV $i$, the relation between two consecutive sensing time slots $\tau_i^{j}$ and $\tau_i^{j+1}$ is given as
\begin{equation}
\sum_{t=\tau_i^{j}}^{\tau_i^{j+1}} \bm{v}_i(t)=\bm{l}_i(\tau_i^{j+1})-\bm{l}_i(\tau_i^{j}), \forall j\in \mathcal{N}_i,
\end{equation}
where $\bm{l}_i(\tau_i^{j})$ and $\bm{l}_i(\tau_i^{j+1})$ are the sensing locations of its $j$th and $j+1$th task. We define the task completion time of UAV $i$ as the number of time slots it costs to complete the sensing and transmission of all its tasks, which can be expressed as
\begin{equation}
T_i=\tau_i^{N_i}+T_{tran,i}^{N_i},
\end{equation}
where $\tau_i^{N_i}$ is the time slot in which it performs the data collection for its last task, and $T_{tran,i}^{N_i}$ is the time that UAV $i$ cost to complete the data transmission for its last task $N_i$.

\section{Sense-and-Send Protocol}
\begin{figure*}[t]
\centerline{\includegraphics[width=6in]{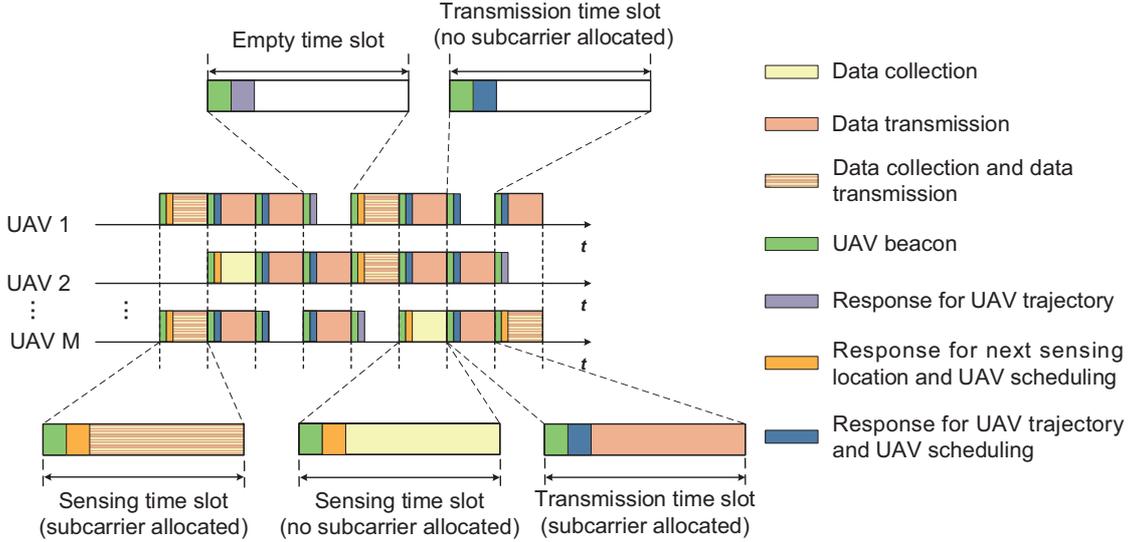}}
\caption{Sense-and-send protocol.}
\label{protocol}
\end{figure*}

In this section, we present the sense-and-send protocol for the UAV cooperation. As illustrated in Fig.~\ref{protocol}, the UAVs perform sensing and data transmission for the tasks in a sequence of time slots. For each UAV, the time slots can be classified into three types: sensing time slot, transmission time slot, and empty time slot. For convenience, we define the location that the UAV performs a sensing task as the \textbf{sensing location}. In each sensing time slot, the UAV collects and transmits data for its current task at the sensing location. In each transmission time slot, the UAV moves along the optimized trajectory and transmits the collected data to the BS. A UAV is in an empty time slot if it has completed data transmission and has not reached the next sensing location. In an empty time slot, the UAV neither collects data nor transmits data, and only moves towards the next sensing location. In the following, we will elaborate on these three types of time slots.

\begin{figure}[h]
\centerline{\includegraphics[width=3in]{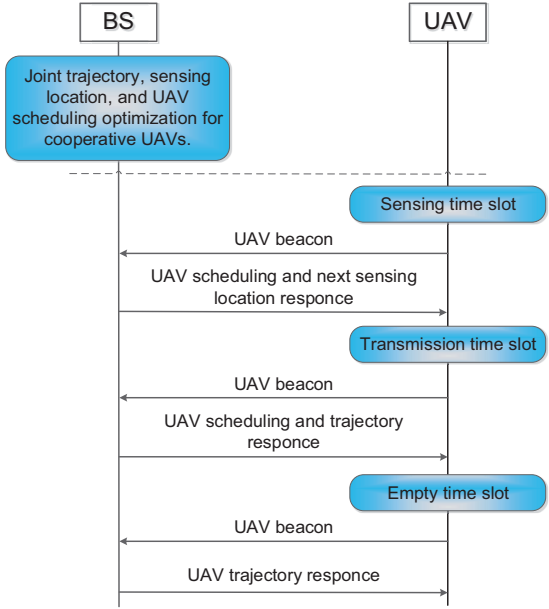}}
\caption{UAV-BS interaction process in the sense-and-send protocol.}
\label{interaction}
\end{figure}

When performing a task, a UAV is in the sensing time slot first, and then switches to the transmission time slot. After several transmission time slots, a UAV may either be in the empty slot or in the next sensing time slot. The BS optimizes the trajectory, sensing location, and UAV scheduling for the cooperative UAVs in advance, and responses the required information to the corresponding UAVs in every time slot. The interaction between the BS and UAV for completing a task is given in Fig.~\ref{interaction}.

\textbf{Sensing Time Slot}: The UAV hovers on the sensing location and performs data collection and transmission in sensing time slot. The UAV first sends UAV beacon to the BS over control channel, which contains the information of its location, the ongoing sensing task, the location of the next sensing task, and transmission request. The BS then informs the UAV of the subcarrier allocation result and sensing location of its next task. Afterwards, the UAV performs data collection until the end of the time slot. The UAV performs data transmission to the BS simultaneously if it is allocated to a subcarrier. For task $j$, the UAV hovers on the sensing location to collect data for only one time slot, with a data collection rate $R_s^j$. If the UAV has not finished data transmission in the sensing time slot, it switches to transmission time slot, otherwise, it switches to an empty time slot.

\textbf{Transmission Time Slot}: When the UAV requires data transmission to the BS, it will operate in the transmission time slot. In the transmission time slot, the UAV first sends UAV beacon to the BS over control channel, which contains the information of transmission request, UAV location, data length to transmit, and the location of the next sensing location. The BS then informs the UAV of its trajectory and UAV scheduling solutions in this time slot. Afterwards, each UAV moves along the optimized trajectory. In the meanwhile, the UAV performs data transmission if it is allocated to a subcarrier. Otherwise, the UAV cannot transmit data in the current time slot and will send data transmission request again in the next transmission time slot. The collected data with respect to task $j$ should be uploaded to the BS before UAV $i$ starts the next sensing task, i.e. $\sum_{t=\tau_i^j+1}^{\tau_i^{j+1}} R_i(t)\geq R_s^j$, where $\tau_i^j$ is the sensing time slot of UAV $i$ for its~$j$th task.

If a UAV does not complete its data transmission in the current time slot, it will occupy another transmission time slot, and request data transmission to the BS again in the next time slot. When a UAV completes data transmission for the current task, it switches to sensing time slot if it has arrived at the sensing location of the next task. Otherwise, it switches to the empty time slot.

\textbf{Empty Time Slot}: A UAV is in the empty time slot when it has completed the data transmission for the current task, and has not arrived at the next sensing location. In empty time slot, the UAV sends UAV beacon that contains its current location and its next sensing location to the BS over control channel. The BS responses the corresponding trajectory to the UAV. The UAV then moves along the optimized trajectory with neither sensing nor transmission in such a time slot. The UAV will switch to sensing time slot when it arrives at the sensing location of the next task.

\textbf{Remark 1:} In each time slot, at most $K$ UAVs can perform data transmission to the BS. When more than $K$ UAVs send transmission request to the BS in one time slot, the UAVs have to share the subcarriers in a time division multiplexing manner.

To describe the signaling cost over the control channels for the proposed protocol, we assume that the UAV beacon contains no more than $\kappa$ messages, and the trajectory, sensing location, and UAV scheduling responce contains at most $\iota$ messages. Therefore, the maximum signaling cost of the network is $M\times(\kappa+\iota)$ in each time slot. The maximum signaling cost is restricted by the number of UAVs, and the signaling of each user costs no more than hundreds of bits~\cite{L2010}. Thus, the signaling cost of the network is tolerable.
\section{Problem Formulation}
In this section, we first formulate the task completion time minimization problem. Afterwards, we decompose it into three sub-problems, and introduce the proposed algorithm that solves the three sub-problems iteratively.
\subsection{Problem Formulation}
Note that the time for completing all the tasks in this network is determined by the maximum task completion time of the UAVs. Let $T_{max}$ be the maximum task completion time of the UAVs in this network, i.e. $T_{max}=\max\{T_i\}, \forall i\in \cal{M}$. To complete all the tasks efficiently, our objective is to minimize the maximum task completion time of the UAVs by optimizing UAV trajectory that consists of speed and direction of the UAV, UAV sensing location, and UAV scheduling. Thus, the problem can be formulated by
\begin{subequations} \label{system_optimization}
\begin{align}\hspace{-2mm}
\mathop{\min}\limits_{\substack{\textbf{$\{\bm{v}_i(t), \bm{l}_i(\tau_i^{j}), \psi_{i}(t)\}$}}} &T_{max},\label{system_1}\\
\textbf{\emph{s.t. }}
&z_i(t)\geq h_{min}, \forall i\in \mathcal{M},\label{system_4}\\
&\|\bm{v}_i(t)\|\leq v_{max}, \forall i\in \mathcal{M},\label{system_5}\\
&\|\bm{v}_i(\tau_i^j)\|=0, \forall i\in \mathcal{M}, j\in \mathcal{N}_i,\label{system_3}\\
&PR_j\geq PR_{th}, \forall j\in \mathcal{N},\label{system_2}\\
&\sum_{t=\tau_i^j+1}^{\tau_i^{j+1}} R_i(t)\geq R_s^j, \forall i\in \mathcal{M}, j\in \mathcal{N}_i,\label{system_7}\\
&\sum_{i=1}^M \psi_{i}(t)\leq K, 1\leq t \leq T_{max},\label{system_8}\\
&\psi_{i}(t)=\{0,1\}.\label{system_9}
\end{align}
\end{subequations}
The altitude and velocity constraints are given by (\ref{system_4}) and~(\ref{system_5}), respectively. (\ref{system_3}) shows that the UAV's velocity is zero when performing sensing, and (\ref{system_2}) implies that the successful sensing probability for each task should be no less than the given threshold $PR_{th}$. Constraint~(\ref{system_7}) explains that the data transmission is required to be completed before the next sensing task, and (\ref{system_8}) is the UAV scheduling constraint.
\subsection{Problem Decomposition}\vspace{-1mm}
Problem (\ref{system_optimization}) contains both continuous variables $\bm{v}_i(t)$ and $\bm{l}_i(\tau_i^{j})$, and binary variable $\psi_{i}(t)$, which is NP-hard. To solve this problem efficiently, we propose an ITSSO algorithm, by solving its three sub-problems: trajectory optimization, sensing location optimization, and UAV scheduling iteratively.

In the trajectory optimization sub-problem, given the sensing locations $\bm{l}_i(\tau_i^{j}), \forall i\in \mathcal{M}, j\in \mathcal{N}_i$ and the UAV scheduling $\psi_{i}(t), \forall i\in \mathcal{M}$, we can observe that different UAVs are independent, and different tasks of a UAV are irrelevant. Therefore, in this sub-problem, the trajectory for a single UAV between two successive tasks can be solved in parallel. Without loss of generality, we study the trajectory for UAV $i$ between its $j$th and $j+1$th task in the rest of this subsection, and the UAV trajectory optimization sub-problem can be written as
\begin{subequations} \label{trajectory subproblem}
\begin{align}\hspace{-2mm}
\mathop{\min}\limits_{\substack{ \textbf{$\bm{v}_i(t)$} }} &(\tau_i^{j+1}-\tau_i^j),\\
\textbf{\emph{s.t. }}
&z_i(t)\geq h_{min}, \forall i\in \mathcal{M},\label{tra-con-1}\\
&\|\bm{v}_i(t)\|\leq v_{max},\label{tra-con-2}\\
&\|\bm{v}_i(\tau_i^j)\|=0,\label{tra-con-3}\\
&\sum_{t=\tau_i^j+1}^{\tau_i^{j+1}} R_i(t)\geq R_s^j.\label{tra-con-4}
\end{align}
\end{subequations}

In the sensing location optimization sub-problem, given the UAV scheduling result and trajectory optimization method, the sensing location optimization sub-problem can be written as\vspace{-2mm}
\begin{subequations} \label{sensing subproblem}
\begin{align}
\mathop{\min}\limits_{\substack{\textbf{$\{\bm{l}_i(\tau_i^{j})\}$} }} &T_{max},\\
\textbf{\emph{s.t. }}
&\|\bm{v}_i(t)\|\leq v_{max}, \forall i\in \mathcal{M},\label{sen-con-1}\\
&\|\bm{v}_i(\tau_i^j)\|=0, \forall i\in \mathcal{M}, j\in \mathcal{N}_i,\label{sen-con-2}\\
&PR_j\geq PR_{th}, \forall j\in \mathcal{N},\label{sen-con-3}\\
&\sum_{t=\tau_i^j+1}^{\tau_i^{j+1}} R_i(t)\geq R_s^j, \forall i\in \mathcal{M}, j\in \mathcal{N}_i.\label{sen-con-4}
\end{align}
\end{subequations}

In the UAV scheduling sub-problem, given the trajectory optimization and sensing location optimization of each UAV, the UAV scheduling sub-problem can be written as\vspace{-2mm}
\begin{subequations} \label{UAV scheduling sub-problem}
\begin{align}
\mathop{\min}\limits_{\substack{\textbf{$\{\psi_{i}(t)\}$}}} &T_{max},\\
\textbf{\emph{s.t. }}
&\sum_{t=\tau_i^j+1}^{\tau_i^{j+1}} R_i(t)\geq R_s^j, \forall i\in \mathcal{M}, j\in \mathcal{N}_i,\label{sch-con-1}\\
&\sum_{i=1}^M \psi_{i}(t)\leq K, 1\leq t \leq T_{max},\label{sch-con-2}\\
&\psi_{i}(t)=\{0,1\}.\label{sch-con-3}
\end{align}
\end{subequations}

\subsection{Iterative Algorithm Description}
In this subsection, we introduce the proposed ITSSO algorithm to solve problem (\ref{system_optimization}), where trajectory optimization, sensing location optimization, and UAV scheduling sub-problems are solved iteratively. We firstly find an initial feasible solution of problem (\ref{system_optimization}) that satisfies all its constraints. In the initial solution, the trajectory, sensing location, and UAV scheduling of this solution are denoted by $\{\bm{v}_i(t)\}^0$, $\{\bm{l}_i(\tau_i^{j})\}^0$, and $\{\psi_{i}(t)\}^0$, respectively. We set the initial sensing location of task $j$ as $(x^n,y^n,h_{min})$ for all the UAVs in $\mathcal{W}_j$. The initial trajectory for UAV $i$ between task $j$ and $j+1$ is set as the line segment between the two sensing locations, with the UAV speed being $v_0$ that satisfies $\frac{v_0}{\bm{l}_i(\tau_i^{j+1})-\bm{l}_i(\tau_i^{j})}\ll \frac{R_s^j}{R_0}$, where $R_0$ is the average transmission rate of UAV $i$ for task $j$. In the initial UAV scheduling the subcarriers are randomly allocated to $K$ UAVs in each time slot.

We then perform iterations of trajectory optimization, sensing location optimization, and UAV scheduling until the completion time for all the tasks converges. In each iteration, the trajectory optimization given in Section~\ref{trajectorysec} is performed firstly with the sensing location optimization and UAV scheduling results given in the last iteration, and the trajectory variables are updated. Next, the sensing location optimization is performed as shown in Section~\ref{sensing}, with the UAV scheduling obtained in the last iteration, and the trajectory optimization results. Afterwards, we perform UAV scheduling as described in Section~\ref{scheduling}, given the trajectory optimization and sensing location optimization results. When an iteration is completed, we compare the completion time for all the tasks obtained in the last two iterations. If the completion time for all the tasks does not decrease with the last iteration, the algorithm terminates and the result is obtained. Otherwise, the ITSSO algorithm continues to the next iteration.

We denote the optimization objective function after the $r$th iteration by $T_{max}\Big(\{\bm{v}_i(t)\}^r, \{\bm{l}_i(\tau_i^{j})\}^r, \{\psi_{i}(t)\}^r\Big)$. In iteration $r$, the trajectory optimization variables $\{\bm{v}_i(t)\}$, the sensing location optimization variables $\{\bm{l}_i(\tau_i^{j})\}$, and the UAV scheduling variables $\{\psi_{i}(t)\}$ are denoted by $\{\bm{v}_i(t)\}^r$, $\{\bm{l}_i(\tau_i^{j})\}^r$, and $\{\psi_{i}(t)\}^r$, respectively. The ITSSO algorithm is summarized in detail as shown in Algorithm \ref{Ite_Alg}.

\begin{algorithm}[htb]\label{Ite_Alg}
\caption{Iterative Trajectory, Sensing, and Scheduling Optimization Algorithm.}
\textbf{Initialization:} Set $r=0$, find an initial solution of problem~(\ref{system_optimization}) that satisfies all its constraints, denote the current trajectory, sensing location, and UAV scheduling by $\{\bm{v}_i(t)\}^0$, $\{\bm{l}_i(\tau_i^{j})\}^0$, and $\{\psi_{i}(t)\}^0$, respectively\;
\While{ $T_{max}\Big(\{\bm{v}_i(t)\}^{r-1}, \{\bm{l}_i(\tau_i^{j})\}^{r-1}, \{\psi_{i}(t)\}^{r-1}\Big)-T_{max}\Big(\{\bm{v}_i(t)\}^r, \{\bm{l}_i(\tau_i^{j})\}^r, \{\psi_{i}(t)\}^r\Big)>0$}{
$r=r+1$\;
Solve the trajectory optimization sub-problem, given $\{\bm{l}_i(\tau_i^{j})\}^{r-1}$ and $\{\psi_{i}(t)\}^{r-1}$\;
Solve the sensing location optimization sub-problem, given $\{\bm{v}_i(t)\}^{r}$ and $\{\psi_{i}(t)\}^{r-1}$\;
Solve the UAV scheduling sub-problem, given $\{\bm{v}_i(t)\}^r$ and $\{\bm{l}_i(\tau_i^{j})\}^r$\;
}
\textbf{Output:}$\{\bm{v}_i(t)\}^r, \{\bm{l}_i(\tau_i^{j})\}^r, \{\psi_{i}(t)\}^r$\;
\end{algorithm}

\section{Iterative Trajectory, Sensing, and Scheduling Optimization Algorithm}
In this section, we first introduce the algorithms that solve the three subproblems (\ref{trajectory subproblem}), (\ref{sensing subproblem}), and (\ref{UAV scheduling sub-problem}), respectively. Afterwards, we analyse the performance of the proposed ITSSO algorithm.
\subsection{Trajectory Optimization}\label{trajectorysec}
In this subsection, we provide a detailed description of the UAV trajectory optimization algorithm (\ref{trajectory subproblem}). Note that we utilize the standard aerial vehicular channel fading model as proposed in~\cite{3GPPR17}, which makes the expression of constraint~(\ref{tra-con-4}) very complicated. Therefore, problem (\ref{trajectory subproblem}) can not be solved with the existing optimization methods. In the following, we optimize the speed and moving direction of the UAVs with a novel algorithm utilizing geometry theorems and extremum principles.

\subsubsection{UAV Speed Optimization}
Assume that the transmission distance from a UAV to the BS is much larger than the UAV velocity, i.e. $d_{i,BS}(t)\gg v_{max}, \forall i\in \mathcal{M}$, we have the following theorem on the optimization of UAV speed.

\textbf{Theorem 1:} The optimal solution can be achieved when the speed of the UAV is $v_{max}$ between $\tau_i^j$ and $\tau_i^{j+1}$, i.e. $\|\bm{v}_i(t)\|=v_{max}, \tau_i^j\leq t\leq \tau_i^{j+1}, \forall i\in \mathcal{M}, j\in \mathcal{N}_i$.
\begin{proof}
See Appendix A.
\end{proof}
According to the proof of Theorem 1, we have the following remark.

\textbf{Remark 2:} With $t'$ being the optimal solution, a trajectory with the length of $t'\times v_{max}$ can be given.

Therefore, we set the UAV speed as $v_{max}$ in the following parts.
\subsubsection{UAV Moving Direction Optimization}
Since the speed of the UAV has been obtained by Theorem 1, we then propose an efficient method to solve the moving direction of UAV $i$ between $\tau_i^j$ and $\tau_i^{j+1}$.

For simplicity, we denote the time between UAV $i$'s $j$th and $j+1$th task by $\delta _i^j=\tau_i^{j+1}-\tau_i^j-1$. Let $[x]$ be the minimum integer that is no smaller than $x$. In problem~(\ref{trajectory subproblem}), the lower bound of $\delta _i^j$ can be expressed as
\begin{equation}
\delta _{i}^{j,lb}=[\frac{\bm{l}_i(\tau_i^{j+1})-\bm{l}_i(\tau_i^{j})}{v_{max}}],
\end{equation}
which corresponds to a line segment trajectory, with $\bm{l}_i(\tau_i^{j+1})-\bm{l}_i(\tau_i^{j})$ being its moving direction. This direction is the solution if constraint~(\ref{tra-con-4}) can be satisfied, i.e. $\sum_{t=\tau_i^j+1}^{\tau_i^{j}+\delta _{i}^{j,lb}} R_i(t)\geq R_s^j$. Otherwise, the UAV has to make a detour to approach the BS for a larger transmission rate, which also leads to a larger task completion time.

\begin{figure}[t]
\centerline{\includegraphics[width=3in]{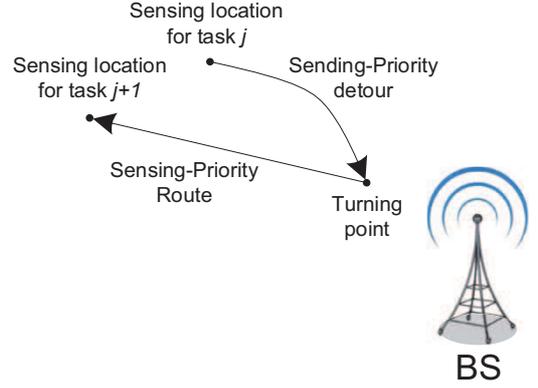}}
\caption{Moving direction optimization.}
\label{trajectory}
\end{figure}

As illustrated in Fig.~\ref{trajectory}, the moving direction of the detoured trajectory contains two parts, namely sending-priority detour and sensing-priority route. In the sending-priority detour part, the UAV detours to the BS for a larger transmission rate, and in the sensing-priority route part, the UAV moves toward its next sensing location with the shortest time.
\begin{itemize}
\item \textbf{Sending-Priority Detour:} In the sending-priority detour part, we maximize the transmission rate of the UAV, so that constraint (\ref{tra-con-4}) can be satisfied with minimum time slots. To achieve the maximum achievable rate, the UAV moves along the direction with the maximum rate ascent velocity, i.e. the gradient of the transmission rate $R_i(t)$, which can be expressed as
\begin{equation}
\nabla R_i(t)=(\frac{\partial R_i(t)}{\partial x}, \frac{\partial R_i(t)}{\partial y}, \frac{\partial R_i(t)}{\partial z}), (\bm{l}_i(t)>h_{min}),
\end{equation}
where the expression of $R_i(t)$ can be derived by equations~(\ref{BS distance}),~(\ref{task distance}),~(\ref{BS receive power})-(\ref{rate}).
In time slot $t$, if the altitude of the trajectory is below the minimum threshold $h_{min}$, the UAV has to adjust its moving direction to
\begin{equation}
\nabla R_i(t)=(\frac{\partial R_i(t)}{\partial x}, \frac{\partial R_i(t)}{\partial y}, 0).
\end{equation}
\item \textbf{Sensing-Priority Route:} We define the endpoint of the sending-priority detour as the \emph{turning point}, denoted by $\bm{l}_i^{tr}(\tau_i^{j})$. In the sensing-priority route part, the UAV moves from the \emph{turning point} to the sensing location of the next task. To minimize the task completion time, the trajectory of the this part is optimized as a line segment, with $\bm{l}_i(\tau_i^{j+1})-\bm{l}_i^{tr}(\tau_i^{j})$ being its moving direction.
\end{itemize}

We denote the time duration of the sending-priority detour and the sensing-priority route by $\delta_i^{j,1}$ and $\delta_i^{j,2}$, respectively. Our target is to find the minimum $\delta_i^{j,1}+\delta_i^{j,2}$ that satisfies constraint (\ref{tra-con-4}). We can observe that a larger $\delta_i^{j,2}$ implies a larger $\delta_i^{j,1}$ since $\delta_i^{j,1}$ is positively related with the detour distance to the BS. Therefore, the minimum $\delta_i^{j,1}+\delta_i^{j,2}$ is achieved with the minimum feasible $\delta_i^{j,1}$. The solution of the moving direction optimization problem is summarized in Algorithm~\ref{Alg1}.

\begin{algorithm}[!thp]\label{Alg1}
\caption{Moving Direction Optimization.}
\textbf{Initialization:} Set moving direction as $\bm{l}_i(\tau_i^{j+1})-\bm{l}_i(\tau_i^{j})$, and $\delta_i^{j,1}=0$\;
\While {Constraint (\ref{tra-con-4}) is not satisfied}{
$\delta_i^{j,1}=\delta_i^{j,1}+1$\;
Optimize the moving direction of sending-priority detour\;
Find the location of the \emph{turning point}\;
Optimize the moving direction of sensing-priority route\;
}
Set the current moving direction as the final solution\;
\end{algorithm}
\subsection{Sensing Location Optimization}\label{sensing}
In this subsection, we propose a method to solve the sensing location optimization subproblem (\ref{sensing subproblem}). It is known that constraint (\ref{sen-con-4}) is non-convex, and thus problem (\ref{sensing subproblem}) can not be solved directly. Since the trajectory between two consecutive sensing locations has been optimized in Section~\ref{trajectorysec}, the UAV trajectory is one-to-one correspondence with the sensing locations.
Note that constraints (\ref{sen-con-1}), (\ref{sen-con-2}), and (\ref{sen-con-4}) can be satisfied with the trajectory optimization method proposed in Section~\ref{trajectorysec} when a sensing location optimization method is given. In the following, we first analyse the properties of the sensing location, and then solve this sub-problem with a local search method.

\textbf{Theorem 2:} For each UAV, the optimal sensing location is collinear with the corresponding \emph{turning point} and the task location\footnote{If the UAV trajectory does not detour to the BS, the start point can be considered as the \emph{turning point}.}.

\begin{figure}[h]
\centering
\includegraphics[width=2in]{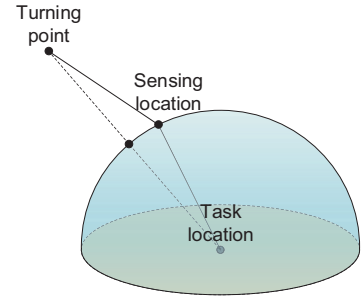}
\caption{Illustration of Theorem 2.}
\label{sensing-location}
\end{figure}
\begin{proof}
To satisfy constraint (\ref{sen-con-3}), we assume that the distance between the sensing location a UAV and the task location should be no more than $d'$ while considering the sensing locations of other UAVs are fixed. As shown in Fig.~\ref{sensing-location}, the feasible solutions of the sensing location is a hemispheroid, with the task location being the center and $d'$ being the radius. We can observe that the sensing location with the shortest trajectory is on the the intersection of the hemispheroid and the line segment from the \emph{turning point} to the task location. Therefore, when the task completion time of a UAV is minimized, the sensing location is collinear with the \emph{turning point} and the task location.
\end{proof}
In the following, we derive the upper bound and lower bound of $\delta _i^j, \forall i\in \mathcal{M}, j\in \mathcal{N}_i$.

\textbf{Theorem 3:} The lower bound of $\delta _i^j$, denoted by $\delta _i^{j,lb}$, is achieved when the sensing location and the \emph{turning point} are overlapped. The upper bound of $\delta _i^j$, denoted by $\delta _i^{j,ub}$, is achieved when the the sensing location of task $j+1$ is $(x^{j+1},y^{j+1},h_{min})$.

\begin{proof}
When the sensing location and the \emph{turning point} are overlapped, the UAV trajectory is the gradient of the transmission rate, which corresponds to the maximum achievable transmission rate. Therefore, constraint (\ref{sen-con-4}) can be satisfied with minimum number of transmission time slots. When the sensing location of task $j+1$ is $(x^{j+1},y^{j+1},h_{min})$, the distance between sensing location and task location is minimized, and the successful sensing probability is maximized. On this condition, the length of the trajectory is maximized, which corresponds to the maximum $\delta _i^j$.
\end{proof}

It is shown that task completion time $T_i$ and successful sensing probability $PR(i,j)$ are negatively correlated. Therefore, a trade-off between the task completion time of UAV $i$ and successful sensing probability is required to minimize the completion time for all the tasks while guaranteeing the the successful sensing probability constraint~(\ref{sen-con-3}).

When solving the sensing location optimization sub-problem, we propose a local search method to reduce the computational complexity. We first give an initial solution that satisfies all the constraints of problem (\ref{sensing subproblem}), and set it as the current solution. The local search method contains iterations of sensing location adjustment. In each iteration, the algorithm is performed task by task. For task $j$, we first reduce the maximum task completion time of the UAVs in set $\mathcal{W}_j$ by one time slot, and then adjust the sensing location of other UAVs in set $\mathcal{W}_j$ to keep constraint (\ref{sen-con-3}) satisfied. The local search method terminates when the completion time for all the tasks can not be reduced by the adjustment of any task.

The detailed process of the sensing location optimization method is elaborated in Algorithm~\ref{Alg2}. When performing the sensing location adjustment for task $j$, without loss of generality, we assume that the maximum task completion time of the UAVs in $\mathcal{W}_j$ is $T_i$. If $\delta _i^j>\delta _i^{j,lb}$, we reduce $\delta _i^j$ by one time slot by adjusting its sensing location. Denote $\mathcal{U}$ by the set of UAVs in $\mathcal{W}_j$ that satisfies $\delta _m^j\leq \delta _m^{j,ub}$, and $T_m\leq T_i-1$, so that the maximum task completion time of the UAVs in $\mathcal{W}_j$ dose not increase. We adjust the sensing locations of the UAVs in set $\mathcal{U}$ to decrease the distances between their sensing locations and the task location until constraint~(\ref{sen-con-3}) is satisfied. Denote the task completion time increment of UAV $m$ by $\Delta t_m$, its sensing location after adjustment is given as $\bm{l}_i(\tau_m^{j})=\bm{l}_i(\tau_m^{j})+\Delta t_m\cdot v_{max}\cdot \frac{\bm{l}_m(\tau_m^{j})-\bm{l}_m^{tr}(\tau_m^{j-1})}{|\bm{l}_m(\tau_m^{j})-\bm{l}_m^{tr}(\tau_m^{j-1})|}, \forall m \in \mathcal{W}_j$. If constraint~(\ref{sen-con-3}) can not be satisfied until $\mathcal{U}$ is empty, the maximum task completion time $T_i$ can not be reduced by adjusting the sensing locations of UAVs in $\mathcal{W}_j$.

\begin{algorithm}[!thp]\label{Alg2}
\caption{Sensing Location Optimization.}
\textbf{Initialization:} Give a set of initial sensing locations $\{\bm{l}_i(\tau_i^{j})\}$\;
\While {The maximum task completion time of any task is reduced in the last iteration}{
\For {$j=1:N$}{
Find the maximum task completion time $T_i$, $\forall i \in\mathcal{W}_j$\;
\If {$\delta _i^j>\delta _i^{j,lb}$}{
$\delta _i^j=\delta _i^j-1$\;
Adjust UAV $i$'s sensing location $\bm{l}_i(\tau_i^{j})$\;
\eIf {Constraint (\ref{sen-con-3}) is satisfied}{
Continue\;
}{
\While {Constraint (\ref{sen-con-3}) is not satisfied}{
\If {$\mathcal{U}=\varnothing$} {Break\;}
\For {$m\in \mathcal{U}$}{
$\delta _m^j=\delta _m^j+1$\;
$\bm{l}_i(\tau_m^{j})=\bm{l}_i(\tau_m^{j})+v_{max}\cdot \frac{\bm{l}_m(\tau_m^{j})-\bm{l}_i^{tr}(\tau_m^{j-1})}{|\bm{l}_m(\tau_m^{j})-\bm{l}_m^{tr}(\tau_m^{j-1})|}$\;
}
}
}
\If {Constraint (\ref{sen-con-3}) cannot be satisfied}{
Recover the adjusted data and continue\;
}
}
}
}
\end{algorithm}

\textbf{Theorem 4:} The iterative sensing location optimization method is convergent.
\begin{proof}
For task $j$, the maximum task completion time of the UAVs in $\mathcal{W}_j$ may reduce or remain unchanged in each iteration. Therefore, the time for completing all the tasks does not increase with the iterations. It is known that the time for completing all the tasks has a lower bound. Therefore, the maximum task completion time can not reduce infinitely, and the iterative sensing location optimization method is convergent.
\end{proof}

\textbf{Theorem 5:} The complexity of the iterative sensing location optimization method is $O(N\times q\times N_i\times M)$.
\begin{proof}
In each iteration of the sensing location optimization method, each of the $N$ tasks is visited for one time, and the total number of UAV sensing for the $N$ tasks is $N\times q$. Therefore, the complexity of each iteration is $N\times q$. The number of iterations is determined by the number of task completion time reduction when performing the local search algorithm. It is known that each UAV has a lower bound of its task completion time, and the task completion time reduction of a UAV is in direct proportion to its task number $N_i$. Therefore, the total number of task completion time reduction is no more than $N_i\times M$, and the number of iteration is no more than $N_i\times M$. In conclusion, the complexity of the iterative sensing location optimization method is $O(N\times q\times N_i\times M)$.
\end{proof}
\subsection{UAV Scheduling}\label{scheduling}
In this subsection, we introduce the UAV scheduling method that solves sub-problem (\ref{UAV scheduling sub-problem}).
Problem (\ref{UAV scheduling sub-problem}) is non-convex since $\psi_{i}(t)$ is a discrete variable, which can not be solved directly. In the following, we propose an efficient method that performs UAV scheduling time slot by time slot.

In each time slot, if the BS receives no more than $K$ transmission requests, each of the request UAV will be allocated with one subcarrier. In the time slot that the BS receives more than $K$ transmission requests, the BS allocates the subcarriers to the $K$ requesting UAVs with maximum task completion time. The task completion time of each UAV is then updated with the change of UAV scheduling in this time slot, and then the BS continues the UAV scheduling of the next time slot. The process of UAV scheduling is shown as Algorithm~\ref{Alg3}.

\begin{algorithm}[!thp]\label{Alg3}
\caption{UAV Scheduling Method.}
\For {$t=1:T_{max}$}{
\eIf {Transmission request number $<K$}{
Allocate a subcarrier to each of the request UAV\;
}{
Allocate a subcarrier to the $K$ requesting UAVs with maximum task completion time\;
}
Update the task completion time of each UAV with the new UAV scheduling\;
}
\end{algorithm}

\subsection{Performance Analysis}\label{PerAna}
In this subsection, we first analyse the performance of the proposed ITSSO algorithm, including its convergence and complexity, and then analyse the system performance of the network.

\emph{1) Convergence:}

\textbf{Theorem 6:} The proposed ITSSO algorithm is convergent.
\begin{proof}
As shown in Section \ref{sensing}, given the trajectory optimization method, the time for completing all the tasks does not increase with the sensing location optimization. In the UAV scheduling, the time for completing all the tasks decreases each time the BS rearranges the subcarriers. Therefore, the time for completing all the tasks does not increase with the iterations of the ITSSO algorithm. It is known that the time for completing all the tasks has a lower bound in such a network, and the objective function can not decrease infinitely. The time for completing all the tasks will converge to a stable value after limited iterations, i.e. the proposed ITSSO algorithm is convergent.
\end{proof}

\emph{2) Complexity:}

\textbf{Theorem 7:} The complexity of the proposed ITSSO algorithm is $O(N^2\times q\times N_i\times M)$.
\begin{proof}
The proposed ITSSO algorithm consists iterations of trajectory optimization, sensing location optimization, and UAV scheduling. In each iteration, the complexity of trajectory optimization is $O(M)$, and the complexity of UAV scheduling is $O(K\times T_{max})=O(K\times N_i)$. The complexity of sensing location optimization is $O(N\times q\times N_i\times M)$, which is proved in Theorem~5. The number of ITSSO algorithm iterations is relevant to the reduction of the time for completing all the tasks, which is in direct proportion to the number of tasks $N$. Therefore, the complexity of the proposed ITSSO algorithm is $O(N\times(M+K\times N_i+N\times q\times N_i\times M))=O(N^2\times q\times N_i\times M)$.
\end{proof}
\emph{3) System Performance Analysis:}
In this part, we analyse the impact of the cooperate UAV number $q$ and the sensing probability threshold $PR_{th}$ on the task completion time of the UAVs in the network.

\textbf{Theorem 8:} The average rate of change of UAV $i$'s task completion time $T_{i}$ to the cooperative UAV number $q$ is
\begin{equation}
\begin{split}
\frac{\Delta T_{max}}{\Delta q}=\frac{(1-PR_{th})^{1/q} \ln(1-PR_{th})}{\lambda(1-(1-PR_{th})^{1/q})q^2}\times\frac{N_i}{v_{max}}.
\end{split}
\end{equation}
\begin{proof}
See Appendix B.
\end{proof}

\textbf{Theorem 9:} The average rate of change of UAV $i$'s task completion time $T_{i}$ to the sensing probability threshold $PR_{th}$ is
\begin{equation}
\begin{split}
\frac{\Delta T_{max}}{\Delta PR_{th}}=\frac{(1-PR_{th})^{1/q-1}}{\lambda q (1-(1-PR_{th})^{1/q})}\times\frac{N_i}{v_{max}}.
\end{split}
\end{equation}
\begin{proof}
See Appendix C.
\end{proof}

In the following, we analyse the dominated factor on the completion time for all the tasks.

\textbf{Theorem 10:} The transmission resource is a dominated factor on the completion time for all the tasks when the network is crowded.
\begin{proof}
We denote the possibility that a UAV in transmission time slot is allocated with a subcarrier by $p_t$, with $p_t\propto K/M$. Therefore, the average time that a UAV finishes data transmission for a task is $\eta\times\frac{M}{K}$, where $\eta$ is a proportionality coefficient. The average time that a UAV cost to finish a task can be given as $\max\{\eta\times\frac{M}{K}, \bar{\delta} ^{lb}\}$, where $\bar{\delta} ^{lb}$ is the average lower bound of the time that a UAV cost to finish a task. Therefore, the transmission resource is the dominated factor on the completion time for all the tasks when $K$ satisfies $\eta\times\frac{M}{K}>\bar{\delta} ^{lb}$, i.e., $K<\frac{\eta\times M}{\bar{\delta} ^{lb}}$.
\end{proof}

We then discuss the impact of sensing task size $R_s^j$ on the completion time for all the tasks in different transmission resource schemes.
\begin{enumerate}
\item \textbf{High Transmission Resource:} In high transmission resource scheme, the most of the UAVs in transmission time slot are allocated with a subcarrier. The impact of sensing task size $R_s^j$ on the completion time for all the tasks is not significant when $R_s^j$ is at a low level, since most of the data transmission tasks can be completed without a detour trajectory. When $R_s^j$ is at a high level, the UAVs are more likely to detour to the BS for data transmission, and $R_s^j$ becomes a dominated factor on the completion time for all the tasks.
\item \textbf{Low Transmission Resource:} In low transmission resource scheme, the subcarriers are occupied by the UAVs in most of the time slots. On this condition, the task completion time of the UAVs are extended, and most of the UAVs detour to the BS for data transmission. A larger sensing task size $R_s^j$ corresponds to a longer sensing-priority detour to the BS i.e. $\frac{\partial T_{max}}{\partial R_s^j}>0$. With the increment of sensing-priority detour, the UAV moves closer to the BS, which improves the average data rate. Therefore, we have $\frac{\partial^2 T_{max}}{\partial (R_s^j)^2}<0$, and the sensing task size $R_s^j$ has a more significant impact on the completion time for all the tasks $T_{max}$ when it is at a low level.
\end{enumerate}

The dominated factor on the completion time for all the tasks is concluded as Table \uppercase\expandafter{\romannumeral1}. The transmission resource $K$ is a dominated factor on the completion time for all the tasks when it is at a low level. The sensing data size $R_s^j$ is a dominated factor of the completion time for all the tasks when $R_s^j$ and $K$ are both at a high level or low level.

\begin{table}[!htbp]
\centering
\caption{Dominated factor on the completion time for all the tasks.}
\begin{tabular}{|c|c|c|c|}
\hline
\multicolumn{2}{|c|}{ \multirow{2}*{} }& \multicolumn{2}{c|}{Transmission Resource $K$}\\
\cline{3-4}
\multicolumn{2}{|c|}{}&High&Low\\
\hline
\multirow{2}*{Sensing Task Size $R_s^j$}&High&$R_s^j$&$K$\\
\cline{2-4}
&Low&$\text{Neither}$&$R_s^j$\& $K$\\
\hline
\end{tabular}
\end{table}
\section{Simulation Results}
In this section, we evaluate the performance of the proposed ITSSO algorithm. The selection of the simulation parameters are based on the existing 3GPP specifications~\cite{3GPPR17} and works~\cite{ZZDS2018}. For comparison, the following schemes are also performed:
\begin{itemize}
\item \textbf{Non-Cooperative (NC) Scheme:} In NC scheme, each task is required to be completed with only one UAV, i.e. $q=1$, and the number of tasks in the network is the same with the proposed ITSSO scheme. The proposed trajectory optimization, sensing location optimization, and UAV scheduling methods are also performed in NC scheme.
\item \textbf{Fixed Sensing Location (FSL) Scheme:} The FSL is given as mentioned in~\cite{ZXZ2018}. In FSL scheme, the sensing locations of the UAVs are given as the location right over the locations of the corresponding tasks, with fixed height $H_{FSL}=50$ m, and the sensing probability constraint~(\ref{system_2}) is not considered in this scheme. The task arrangement for each UAV in FSL scheme is the same with the proposed ITSSO scheme, and the proposed trajectory optimization and UAV scheduling methods are utilized in FSL scheme.
\end{itemize}
For simulation setup, the initial location of the UAVs are randomly and uniformly distributed in an 3-dimension area of 500 m $\times$ 500 m $\times$ 100 m, and the tasks are uniformly distributed on the ground of this area. We assume that the number of tasks for different UAVs are equal, i.e. $N_i=N_j=\frac{N\times q}{M}, \forall i,j \in \mathcal{M}$, and the task arrangement for each UAV is given randomly. The data collection rate for different tasks are fixed, denoted by $R_s^j=R_s, \forall j\in \mathcal{N}$. The values of $N$, $q$, and $m$ are given in each figure. All curves are generated by over 10000 instances of the proposed algorithm. The simulation parameters are listed in Table \uppercase\expandafter{\romannumeral2}.

\renewcommand\arraystretch{2}

\begin{table}[!t]
\centering
\caption{Simulation Parameters}
\begin{tabular}{|c||c|}
\hline
\textbf{Parameter} & \textbf{Value}\\
\hline
\hline
BS height $H$ & 25 m\\
\hline
Carrier frequency $f_c$& 2 GHz\\
\hline
Number of subcarriers $K$ & 10\\
\hline
Bandwidth of each subcarrier $W_B$ & 1MHz\\
\hline
Sensing task size $R_s$ & 20Mbps\\
\hline
Noise variance $\sigma^2$ & -96 dBm\\
\hline
UAV transmit power $P_U$ & 23dBm\\
\hline
Maximum UAV velocity $v_{max}$ & 50 m/s\\
\hline
Minimum UAV altitude $h_{min}$ & 10 m\\
\hline
Sensing performance parameter $\lambda$ & 0.01\\
\hline
Sensing probability threshold $PR_{th}$ & 0.9\\
\hline
\end{tabular}
\end{table}

\begin{figure}[ht]
\centerline{\includegraphics[width=3.2in]{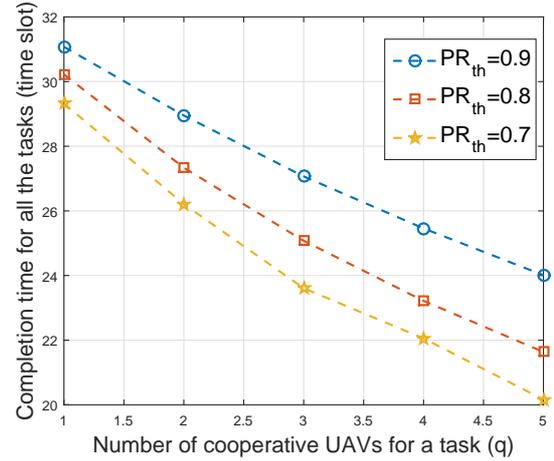}}\vspace{-3mm}
\caption{Number of cooperative UAVs for a task vs. Completion time for all the tasks ($N$=20, $N_i$=4).}\label{Fig4}\vspace{-3mm}
\end{figure}
Fig.~\ref{Fig4} depicts the completion time for all the tasks $T_{max}$ v.s. the number of cooperative UAVs $q$ for each task. The number of tasks in the network is set as 20, and each UAV is arranged to complete 4 tasks. The number of UAVs $M$ varies with variable $q$, satisfying $M\times N_i=N\times q$. It is shown that the completion time for all the tasks decreases with the increment of cooperative UAVs for a task. The reason is that the average distance between the UAV and the task when performing data collection increases with a larger cooperative UAV number, which corresponds to a shorter moving distance. The slopes of the curves decrease with the increment of $q$, which satisfies the theoretical results given in Theorem 8. The completion time for all the tasks decreases for about 8\% when we change the sensing probability threshold from 0.9 to 0.8, and it further decreases about 6\% if the threshold is reduced to 0.7.

\begin{figure}[ht]
\centerline{\includegraphics[width=3.2in]{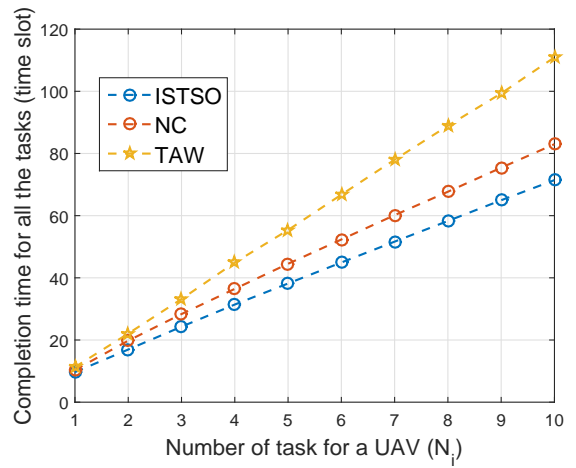}}\vspace{-3mm}
\caption{Number of tasks for a UAV vs. Completion time for all the tasks ($M$=20, $q$=4).}\label{Fig1}\vspace{-3mm}
\end{figure}
Fig.~\ref{Fig1} shows the completion time for all the tasks $T_{max}$ v.s. the number of tasks for a UAV $N_i$. In the ITSSO and FSL schemes, we set the number of UAVs as 20, and the number of cooperative UAVs for a task as 4. In the NC scheme, the number of UAVs is also set as 20, and each task is only performed with one UAV. The completion time for all the tasks increases linearly with the number of tasks for a UAV. The slope of the ITSSO scheme is about 15\% lower than that of the NC scheme due to a shorter average UAV moving distance. The completion time for all the tasks of the FSL scheme is over 50\% larger than the ITSSO scheme due to the lack of sensing location optimization.

\begin{figure}[ht]
\centerline{\includegraphics[width=3.2in]{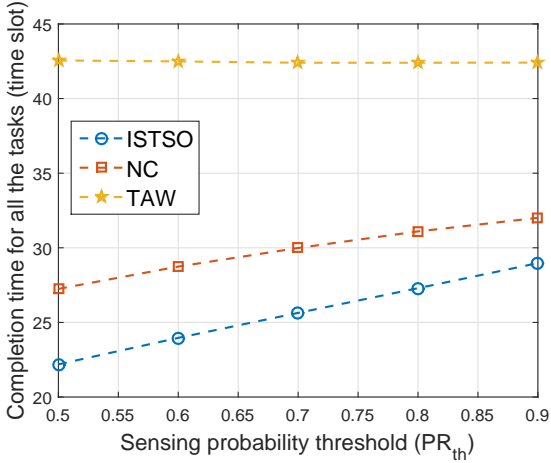}}\vspace{-3mm}
\caption{Sensing probability threshold vs. Completion time for all the tasks ($N$=20, $N_i$=4, $q$=4).}\label{Fig3}\vspace{-3mm}
\end{figure}
In Fig.~\ref{Fig3}, we plot the relation between the sensing probability threshold $PR_{th}$ and the completion time for all the tasks $T_{max}$. Here, we set the number of tasks as 20, and the number of tasks for a UAV as 4. In the ITSSO and FSL schemes, the number of cooperative UAVs for a task is set as 4. In the NC scheme, each task is performed by only one UAV. We can observe that the completion time for all the tasks of the ITSSO and NC scheme increases with the sensing probability threshold due to the change of the sensing locations. The completion time for all the tasks in the ITSSO scheme increases from 22 to 29 when the sensing probability threshold increases from 0.5 to 0.9, and the completion time for all the tasks in the NC scheme increases from 27 to 32 in the same range of sensing probability threshold. The slope is consistent with the theoretical results given in Theorem 9. The completion time for all the tasks in the FSL scheme is around 42, since the sensing locations are fixed in this scheme.

\begin{figure}[ht]
\centerline{\includegraphics[width=3.2in]{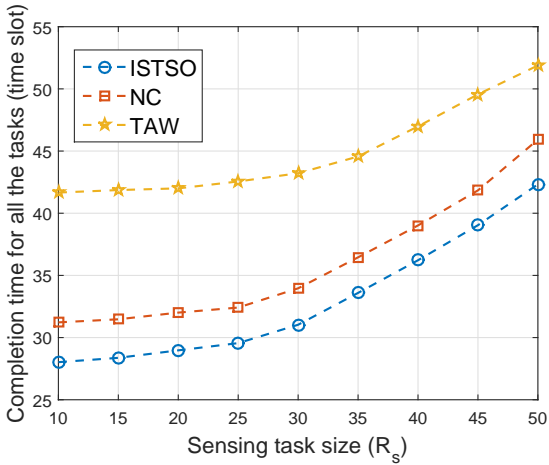}}\vspace{-3mm}
\caption{Sensing task size vs. Completion time for all the tasks ($N$=20, $N_i$=4, $q$=4).}\label{Fig5}\vspace{-3mm}
\end{figure}
Fig.~\ref{Fig5} shows the completion time for all the tasks as a function of the sensing task size $R_s$. We set the number of tasks as 20, and the number of UAVs as 10. In the ITSSO and FSL scheme, the number of cooperative UAVs for a task is set as 4. In the NC scheme, each task is performed by only one UAV. For the ITSSO scheme, the completion time for all the tasks is around 28-30 when $R_s\leq 25$ Mbps, where most of the data transmission tasks can be completed without a detour trajectory. The completion time for all the tasks starts to increase significantly when $R_s> 25$ Mbps, since the UAVs are more likely to detour to the BS for data transmission. Note that the increment of the completion time for all the tasks is mainly caused by the trajectory detouring. Therefore, the completion time for all the tasks in all the three schemes increases with the sensing task size, and the difference among the three schemes decreases with a larger $R_s$.

\begin{figure}[ht]
\centerline{\includegraphics[width=3.2in]{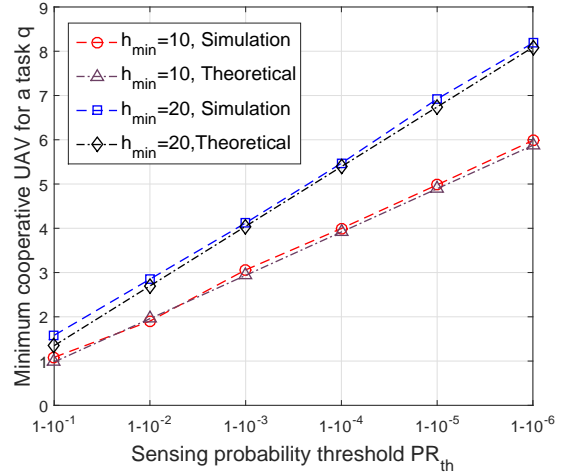}}\vspace{-3mm}
\caption{Sensing probability threshold vs. Minimum number of cooperative UAVs for a task.}\label{Fig2}\vspace{-3mm}
\end{figure}
In Fig.~\ref{Fig2}, we study the minimum number of cooperative UAVs required for completing a task with different sensing probability thresholds. Given a high sensing probability requirement, UAV cooperation is necessary for the UAVs to complete a sensing task. We can observe that the minimum number of UAVs required for a task is logarithmically related to the sensing probability threshold, which is consistent with the theoretical results in (\ref{UAV sensing probability}) and (\ref{cooperate sensing}). The difference between the simulation result and the theoretical result is less than 0.2 within the simulation range. When we set $PR_{th}=1-10^{-1}$, the minimum number of UAVs required for a task is 1. When the sensing probability threshold is set as $PR_{th}=1-10^{-6}$, at least 6 UAVs are required to complete a task. The required number of cooperative UAVs increases for about 50\% when we adjust the minimum UAV altitude $h_{min}$ from 10 m to 20~m.

\begin{figure}[ht]
\centerline{\includegraphics[width=3.2in]{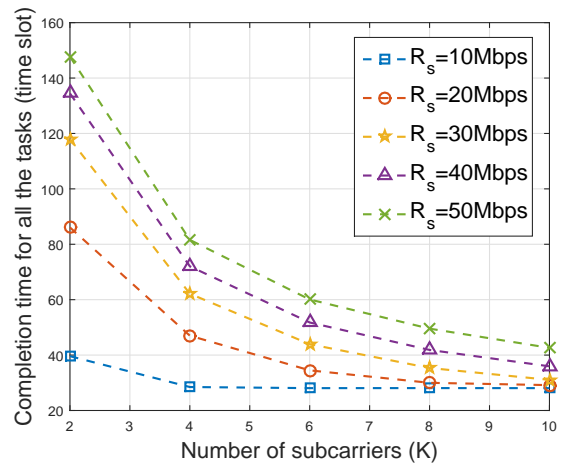}}\vspace{-3mm}
\caption{Number of subcarriers vs. Completion time for all the tasks ($q=4$, $M=20$).}\label{Fig6}\vspace{-3mm}
\end{figure}
Fig.~\ref{Fig6} illustrates the impact of subcarrier number $K$ and task size $R_s$ on the completion time for all the tasks. The number of UAV is set as $M=20$, and number of cooperative UAV for a task is set as $q=4$. It is shown that the completion time for all the tasks is strongly affected by the task size $R_s$ with $R_s\leq 30$~Mbps, and the marginal impact of $R_s$ decreases when $R_s> 30$~Mbps. Given a fixed $R_s$, the completion time for all the tasks decreases rapidly with the number of subcarriers with $K\leq4$. The ratio decreases until convergence with the increment of the number of subcarriers. The simulation curves satisfies the analysis of the dominated factor of the completion time for all the tasks given in Section~\ref{PerAna}.

\section{Conclusions}
In this paper, we studied a single cell UAV sensing network where multiple cooperative UAVs perform sensing and transmission. We first proposed a sense-and-send protocol to facilitate the cooperation, and formulated a joint trajectory, sensing location, and UAV scheduling optimization problem to minimize the completion time for all the tasks. To solve the NP-hard problem, we decoupled it into three sub-problems: trajectory optimization, sensing location optimization, and UAV scheduling, and proposed an iterative algorithm to solve it. We then analyzed the system performance, from which we can infer that the UAV cooperation reduces the completion time for all the tasks, and the marginal gain becomes smaller with the increment of cooperative UAV number. Simulation results showed that the completion time for all the tasks in the proposed ITSSO scheme is 15\% less than the NC scheme, and over 50\% less than the FSL scheme. The transmission resource is a dominated factor on the completion time for all the tasks when it is at a low level. The sensing data size is a dominated factor of the completion time for all the tasks when the size of the sensing task and the transmission resource are both at a high level, or both at a low level.

\begin{appendices}
\section{Proof of Theorem 1}
\begin{proof}
We assume that in the optimal trajectory, there exists a time slot $t_0$, in which the speed of the UAV is $\|\bm{v}_i(t_0)\|=v'$, with $v'<v_{max}$. In the following, we will prove that there exists a solution with $\|\bm{v}_i(t_0)\|=v_{max}$, whose performance is no worse than the one with $\|\bm{v}_i(t_0)\|=v'$.

Given the location of $\bm{l}_i(t_0-1)$ and $\bm{l}_i(t_0+1)$, the possible location of $\bm{l}_i(t_0)$ is shown as the two red points in Fig. \ref{appA-1}. When we set $\|\bm{v}_i(t_0)\|=v_{max}$, the possible location of the UAV in time slot $t_0$ moves to the blue points. If the BS is on the right side of the polyline, at least one blue point is nearer to the BS than both of the red points. Therefore, setting $\|\bm{v}_i(t_0)\|=v_{max}$ can improve the transmission rate in time slot $t_0$ when the BS is on the right side of the polyline.

Similarly, we analyse the possible location of $\bm{l}_i(t_0-1)$ in Fig. \ref{appA-2}. Given the location of $\bm{l}_i(t_0-2)$ and $\bm{l}_i(t_0)$, the possible location of $\bm{l}_i(t_0-1)$ is shown as the two red points with $\|\bm{v}_i(t_0)\|=v'$. When we set $\|\bm{v}_i(t_0)\|=v_{max}$, the possible location of the UAV in time slot $t_0-1$ moves to the blue points. Similarly, we get a polyline, on the left side of which at least one blue point is nearer to the BS than both of the red points. The transmission rate in time slot $t_0-1$ can be improved when the BS is on the left side of the new polyline.

It can be easily proved that the two polylines are not parallel, and a quadrangle is obtained with the two polylines. Since the transmission distance is much larger than the UAV velocity, i.e. $d_{i,BS}(t)\gg v_{max}$, the BS is outside of the quadrangle. Therefore, the transmission rate of at least one time slot can be improved when $\|\bm{v}_i(t)\|$ is changed from $v'$ to $v_{max}$. A larger transmission rate reduces the number of transmission time slots, and thus, the task completion time is no more than the solution with $\|\bm{v}_i(t_0)\|=v'$.
\begin{figure}
\centering
\subfigure[]{
\begin{minipage}[b]{2.6in}\label{appA-1}\vspace{-3mm}
\includegraphics[width=2.4in]{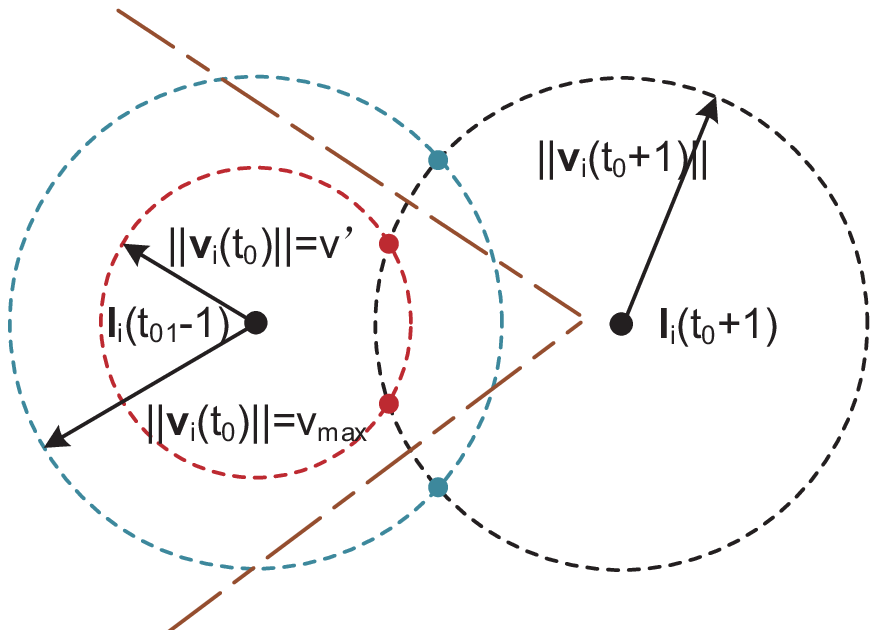}\vspace{-3mm}
\end{minipage}
}
\subfigure[]{
\begin{minipage}[b]{2.6in}\label{appA-2}\vspace{-3mm}
\includegraphics[width=2.4in]{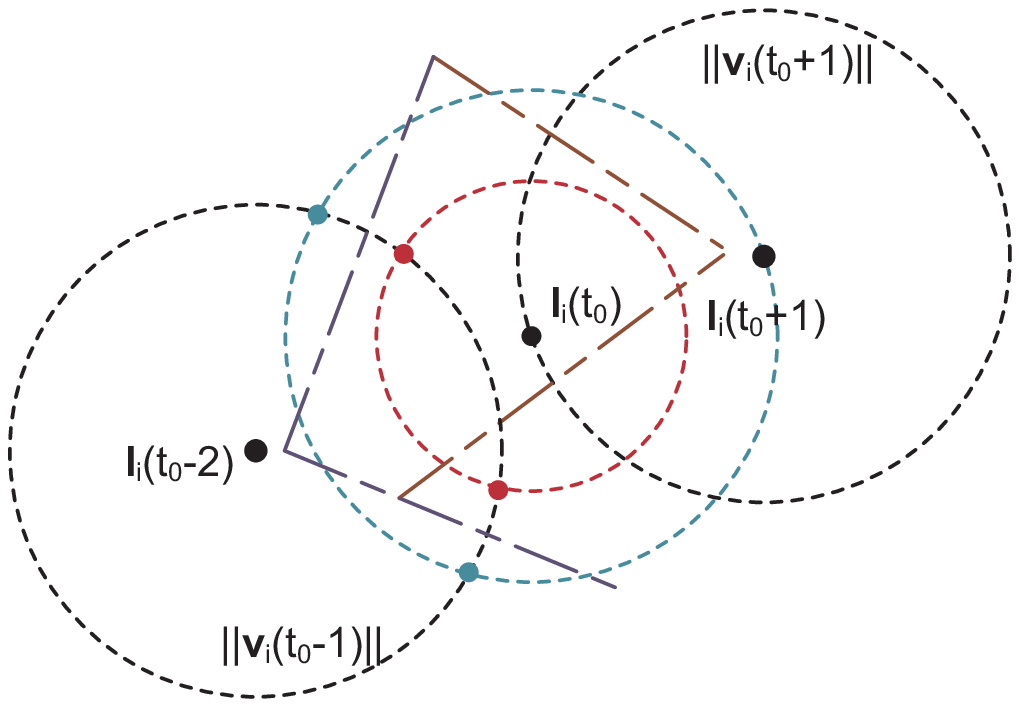}\vspace{-3mm}
\end{minipage}
}\vspace{-3mm}
 \caption{Proof of Theorem 1.} \label{appA}\vspace{-3mm}
\end{figure}
\end{proof}

\section{Proof of Theorem 8}
\begin{proof}
According to equations (\ref{UAV sensing probability}) and (\ref{cooperate sensing}), a larger cooperative UAV number $q$ requires lower successful sensing probability for each UAV. Given the sensing probability threshold $PR_{th}$, when a UAV performs data collection, the distance between the UAV and the sensing task can be longer with a larger $q$. When considering the average change of rate, we assume that the distance between every UAV and the sensing task are the same when performing data collection, and the sensing probability of each task equals the sensing probability threshold $PR_{th}$, i.e.
\begin{equation}\label{d equal}
d_{i,j}(\tau_i^j)=d_0, \forall i,m\in \mathcal{W}_j,
\end{equation}
\begin{equation}\label{PR equal}
PR_j=PR_{th}, \forall j\in \mathcal{N}.
\end{equation}
When substituting (\ref{d equal}) and (\ref{PR equal}) into (\ref{UAV sensing probability}) and (\ref{cooperate sensing}), we have
\begin{equation}\label{equal}
PR_{th}=1-(1-e^{-\lambda d_0})^q.
\end{equation}
The average rate of change of sensing distance to the cooperate UAV number can be achieved with the derivation of $q$ in equation (\ref{equal}), which is shown as
\begin{equation}
\frac{\Delta d_0}{\Delta q}=-\frac{(1-PR_{th})^{1/q} \ln(1-PR_{th})}{\lambda(1-(1-PR_{th})^{1/q})q^2}.
\end{equation}
For each UAV, the increment of sensing distance equals to the decrement of moving distance. Given the UAV speed $v_{max}$, the average rate of change of time for each task to the cooperate UAV number is $\frac{\Delta \delta_i^j}{\Delta q}=-\frac{\Delta d_0}{v_{max}\Delta q}$. For UAV $i$, the average rate of change of its task completion time to the cooperate UAV number is $\frac{\Delta T_i}{\Delta q}=-\frac{N_i \Delta d_0}{v_{max} \Delta q}=\frac{(1-PR_{th})^{1/q} \ln(1-PR_{th})}{\lambda(1-(1-PR_{th})^{1/q})q^2}\times\frac{N_i}{v_{max}}$.
\end{proof}
\vspace{-4mm}
\section{Proof of Theorem 9}
\begin{proof}
Similar with the proof of Theorem 8, we assume that the distance between every UAV and the sensing task are the same when performing data collection, and the sensing probability of each task equals the sensing probability threshold $PR_{th}$. Therefore, equations (\ref{d equal}), (\ref{PR equal}), and (\ref{equal}) are also satisfied in the prove of Theorem 9. The average rate of change of sensing distance for one task to the sensing probability threshold can be achieved with the derivation of $PR_{th}$ in equation (\ref{equal}), which is shown as
\begin{equation}
\frac{\Delta d_0}{\Delta PR_{th}}=-\frac{(1-PR_{th})^{1/q-1}}{\lambda q (1-(1-PR_{th})^{1/q})}.
\end{equation}
For each UAV, the increment of sensing distance equals to the decrement of moving distance. Given the UAV speed $v_{max}$, the average rate of change of time for each task to the sensing probability threshold is $\frac{\Delta \delta_i^j}{\Delta PR_{th}}=-\frac{\Delta d_0}{v_{max}\Delta PR_{th}}$. For UAV $i$, the average rate of change of its task completion time to the sensing probability threshold is $\frac{\Delta T_i}{\Delta PR_{th}}=-\frac{N_i \Delta d_0}{v_{max} \Delta PR_{th}}=\frac{(1-PR_{th})^{1/q-1}}{\lambda q (1-(1-PR_{th})^{1/q})}\times\frac{N_i}{v_{max}}$.
\end{proof}
\end{appendices}

\end{document}